\documentclass[useAMS,usenatbib,usegraphicx]{mn2e}

\title[Dark matter distribution in M\,31]{Visible and dark matter in M\,31 -
II. A dynamical model and dark matter density distribution}
\author[E. Tempel, A. Tamm and P. Tenjes]{E. Tempel$^{1,2}$\thanks{E-mail:
elmo@aai.ee;
atamm@ut.ee; peeter.tenjes@ut.ee}, A. Tamm$^{1}$ and P. Tenjes$^{1,2}$\\
$^{1}$Tartu Observatory, 61602 T\~oravere, Estonia\\
$^{2}$Institute of Theoretical Physics, Tartu University, T\"ahe 4,
51050 Tartu, Estonia}

\begin{document}
\date{Accepted 2008 Month 00, Received 2007 Month 00.}
\pagerange{\pageref{firstpage}--\pageref{lastpage}} \pubyear{2008}
\maketitle
\label{firstpage}
\begin{abstract}
In the present paper we derive the density distribution of dark
matter (DM) in a well-observed nearby disc galaxy, the Andromeda
galaxy. From photometrical and chemical evolution models constructed
in the first part of the study \citep*[][hereafter Paper~I]{tamm:08}
we can calculate the mass distribution of visible components
(the bulge, the disc, the stellar halo, the outer diffuse
stellar halo). In the dynamical model we calculate stellar rotation
velocities along the major axis and velocity dispersions along the
major, minor and intermediate axes of the galaxy assuming
triaxial velocity dispersion ellipsoid. Comparing the calculated
values with the collected observational data, we find the amount of
DM, which must be added to reach an agreement with the observed
rotation and dispersion data.

We conclude that within the uncertainties, the DM distributions by
Moore, Burkert, Navarro, Frenk \& White (NFW) and the
Einasto fit with observations nearly at all distances. The NFW and
Einasto density distributions give the best fit with observations.

The total mass of M\,31 with the NFW DM distribution is
$1.19\cdot 10^{12}\rmn{M_{\sun}}$, the ratio of the DM mass to the
visible mass is 10.0. For the Einasto DM distribution, these values
are $1.28\cdot 10^{12}\rmn{M_{\sun}}$ and 10.8. The ratio of the DM
mass to the visible mass inside the Holmberg radius is 1.75 for the
NFW and the Einasto distributions. For different cuspy DM
distributions, the virial mass is in a range (6.9--7.9)$\cdot10^{11}
\rmn{M_{\sun}}$ and the virial radius is $\sim150\,\rmn{kpc}$. The DM
mean densities inside 10\,pc for cusped models are 33 and
16$\,\rmn{M_{\sun}pc^{-3}}$ for the NFW and the Einasto profiles,
respectively. For the cored Burkert profile, this value is
0.06$\,\rmn{M_{\sun}pc^{-3}}$.
\end{abstract}

\section{Introduction}

Models of hierarchical merging of cold dark matter
describe rather well the observed general properties of galaxies,
large-scale structure of clusters, superclusters and their network.

On the other hand, the observed number of dwarf galaxies
seems to be too small in comparison with cold dark matter models. But by far the
most commonly referred contradiction is related to disc galaxies.
Cosmological simulations generate dark matter (DM) haloes with
central density cusps of $\rho \sim r^{-1}$ or steeper
\citep*[e.g.][]{navarro:97,moore:99}. Observations of dwarf and low
surface brightness disc galaxies have usually shown that shallow
central density profiles fit the data better than cuspy profiles
\citep*{moore:94,burkert:95,blaisouellette:01,borriello:01,deblok:01,
deblok:02,salucci:03,weldrake:03,simon:05,
zackrisson:06,kassin:06,gentile:04,gentile:07,valenzuela:07}. Observational data
are consistent with cuspy density profiles only for a few objects
\citep[e.g.][]{swaters:03,deblok:05,simon:05}.

Discrepancy between the predicted cuspy DM density profiles
and the observed rotation curves with shallow profiles has been
explained with limited resolution of rotation curves and
non-circular motions of gas and with the statement that the actual
contribution of visible stellar matter is poorly known.

The aim of the present paper is to derive the density distribution
of dark matter in a nearby disc galaxy, the Andromeda galaxy. The galaxy
M\,31 was selected because (1)~photometrical and kinematical
(rotation, dispersions) data are measured with sufficiently high
resolution in order to study the bulge region; (2)~velocity
dispersions have been measured also outside the galactic apparent
major axis; in addition to stellar kinematics, the kinematics of
planetary nebulae (PN) is known; (3)~direct measurements of
metallicities allow to constrain the mass-to-light~($M/L$) ratios of
visible matter; (4)~independent estimates of the mass distribution
on large scales are available (globular clusters (GC), satellites,
stream, kinematics of the Milky~Way\,+\,M\,31 system).

In the first part of the study (Paper~I), we constructed the
photometrical model of M\,31 stellar populations on the basis of
surface brightness profiles in {\it U\/}, {\it B\/}, {\it V\/}, {\it
R\/}, {\it I\/} and {\it L\/} colours \citep*[see also][]{tenjes:94a}.
The derived photometrical model gives us parameters of galactic
components, colour indices among them. From independent spectral
observations, the metallicity of the stellar content is available.
The colour indices and the metallicity of each component were
interpreted with the help of chemical evolution models to calculate
the ages and $M/L$-ratios of the components. In total, the output of
the first paper were density distribution parameters, ages and
$M/L$-ratios of galactic components.

In the present paper, we apply the results of Paper~I and construct
a mass distribution model of M\,31 consistent with the measured
kinematics. Stellar components and their $M/L$'s together with the
photometrical model give us the mass distribution of visible matter.
Calculating the rotation velocities and velocity dispersions of
visible matter with the help of the dynamical model
\citep[see][]{tempel:06} the amount of DM can be found, which must be
added to reach an agreement with the observed rotation and
dispersion data.

In principle, to derive the DM density distribution, it is not
necessary to take into account stellar velocity dispersions. It is
sufficient to compare the gas rotation curve with the calculated
derivatives of the gravitational potential. This simple and
straightforward method is frequently used. Of course, before
calculations the gas rotation curve must be corrected for possible
expansion velocities, non-circular motions and dispersions
\citep[see][]{gentile:07}. However, one of the most uncertain
aspects of this method is the $M/L$ of the visible matter. Chemical
evolution models involve several insufficiently constrained
parameters. It is possible to decrease degeneracy by using 3--5
different colour indices, but due to measurement uncertainties,
colour indices are often controversial. The comparison of the
results with additional and independent observations (stellar
rotation curve, velocity dispersions along several slit-positions)
allows us to constrain the distributions of visible matter and
thereafter DM.

In the present model, we assume that a galaxy is a superposition of
ellipsoids of rotation with different flattening. We assume that
stellar populations have both rotation and dispersion components.
The velocity dispersion ellipsoid of visible matter is triaxial. DM
has spherical density distribution and is collision-free.

\begin{figure}
\includegraphics[width=84mm]{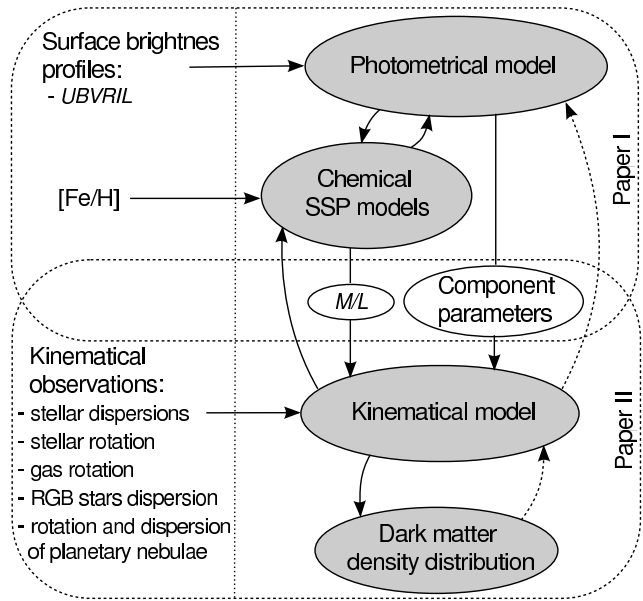}
\caption{The principle scheme of the construction of the M\,31 model
in the framework of Paper~I and Paper~II (the present paper). The
arrows show dependencies between different model construction
stages. The dashed arrows represent weaker impact.} \label{scheme}
\end{figure}

The general scheme of Paper~I and Paper~II (present paper) is given
in Fig.~\ref{scheme}. The photometrical model and the chemical
evolution model constitute the first part of the study and depend
very strongly on each other. The output of the first part (component
parameters and $M/L$-ratios) is the input to the second part of the
study. At the second stage, we use the kinematical model and narrow
down the output of the first part. Finally, we get a self-consistent
model and derive the DM density distribution, using a possibly wide
variety of observational data.

In Section~\ref{sec:2}, we describe the observational data which we have used.
Kinematical data consist of gas rotation, stellar rotation and
dispersions, rotation velocities and dispersions of PN and velocity
dispersions of individual red giant branch (RGB) stars. In
Section~\ref{sec:3}, the dynamical model and an algorithm for calculating the
DM distribution are described. In Sections~\ref{sec:4} and~\ref{sect_dm},
visible and DM
density distributions are derived and compared with available
observational data. In Section~\ref{sec:6}, the results are discussed.

The general parameters of M\,31 used in the present paper are: the
system velocity is \hbox{-300\,km\,s$^{-1}$}
\citep{devaucouleurs:91,vandenbergh:00}, the galaxy inclination angle is 77.5\degr
\citep{walterbos:88,devaucouleurs:91}, the major axis position angle is 38.1\degr
\citep{walterbos:87,ferguson:02} and the distance has been taken 785\,kpc
\citep{mcconnachie:05}, corresponding to the scale \hbox{1\,arcmin =
228\,pc}. Throughout the paper we use the $\Lambda$CDM
cosmological model with $\Omega_{\rmn{m}}=0.3$ and
$H_0=75\,\rmn{km\,s^{-1}Mpc^{-1}}$.

\section{Observational data}\label{sec:2}

In Paper~I we described photometrical and chemical composition data
used to construct the photometrical model. In the present section,
we describe kinematical data, used for the construction of the
dynamical model. From a large variety of kinematical information, we
present only the observations, that can be compared with our model
output.

\subsection{H\,{\sevensize\bf I} and H\,{\sevensize\bf II} observations}

\begin{figure}
\includegraphics[width=84mm]{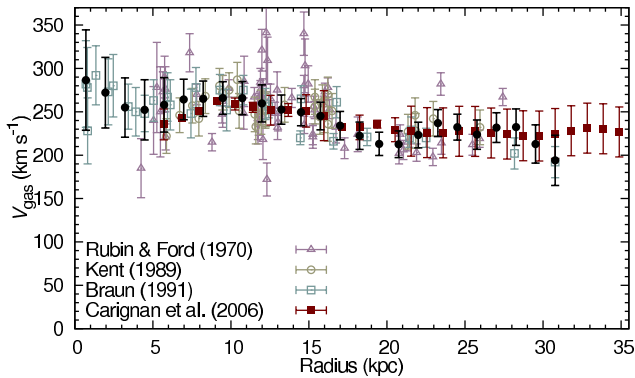}
\caption{Circular velocity, derived from H\,{\sc i} and H\,{\sc ii}
observations. The filled circles are averaged rotational velocities
obtained using the data from earlier observations (faint points).
The filled squares are recent H\,{\sc i} observations by
\citet{carignan:06}. See the text for more information.
}\label{bootgasvel}
\end{figure}

We constructed the gas rotation curve of M\,31 on the basis of
H\,{\sc ii} observations by \citet{rubin:70} and H\,{\sc i} observations
by \citet{kent:89} and \citet{braun:91}. Observational data points from these
studies were averaged, and the resulting rotation velocities are
presented by filled circles in Fig.~\ref{bootgasvel}. To calculate
the error-bars, a bootstrap method is used, combining the bootstrap
errors with observational errors.

In outer parts of M\,31, an extended H\,{\sc i} rotation curve has
recently been measured by \citet{carignan:06}. They measured the rotation
curve outside \hbox{90\,arcmin}, between 20 and 35\,kpc, and
recomputed also rotation velocities, using earlier H\,{\sc i} data
by \citet{unwin:83}. Rotational velocities by \citet{carignan:06} are showen
by filled squares in Fig.~\ref{bootgasvel}. The derived rotation
curve represents gas kinematics from 1--35\,kpc.

\subsection{Stellar velocity and dispersion observations}

\begin{table}
\caption{References for rotation and dispersion data along the major
and the minor axis.} \label{table_kin}
\begin{tabular}{@{}llll}
\hline
Reference               & $R_\rmn{max}$ [kpc] & $z_\rmn{max}$ [kpc] \\
\hline
\citet{mcelroy:83}      & 2.27 & 1.98 \\
\citet{kormendy:88a}      & 0.185& -- \\
\citet{vandermarel:94}& 0.234& 0.16 \\
\citet{kormendy:99}     & 0.18 & -- \\
\hline
\end{tabular}
\end{table}
\begin{figure}
\includegraphics[width=84mm]{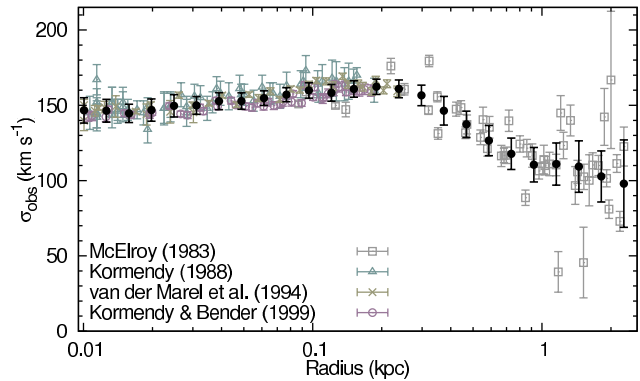}
\caption{Stellar dispersion observations. The faint points
correspond to observations from different authors. The filled
circles denote the averaged profile.}\label{bootdmaj}
\end{figure}
\begin{figure}
\includegraphics[width=84mm]{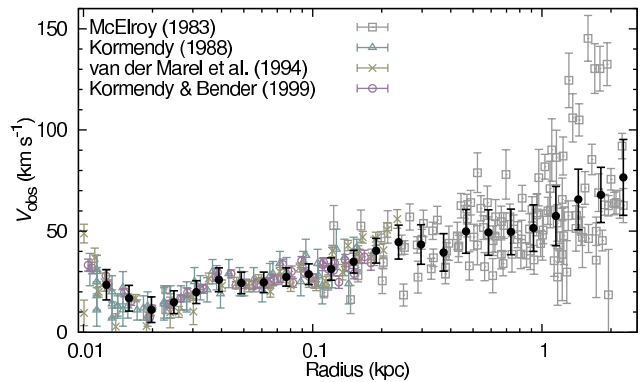}
\caption{Stellar rotational velocities. The faint points are
observations from different authors. The filled circles denote the
averaged profile.}\label{bootvmaj}
\end{figure}

The intrinsic structure of the nucleus of M\,31 is not clear -- it
may have a ringlike and/or a double or even a triple structure
\citep*{lauer:93,gerssen:95,peiris:03,salow:04,bender:05}. In The context of the present paper,
the contribution of the nucleus is unimportant, we look at regions
outside 0.01\,kpc and we exclude the nucleus from our model.

Stellar rotation velocities and velocity dispersions in the central
regions of M\,31 have been measured by several authors \citep[see
references in][]{tenjes:94a}. To model the contribution of visible and DM
to the total mass distribution, it is useful to possess kinematical
data outside the galactic apparent major axis and at largest
distance intervals \citep[see][]{tempel:06}. References to stellar
kinematical data used in the present paper are presented in
Table~\ref{table_kin} which also gives maximal distances along the
major and the minor axis. \citet{mcelroy:83} measured stellar rotation and
dispersions up to the distances of 2.3\,kpc from the centre along
12~cuts (at every $15\degr$) that penetrate the galaxy centre. None
of them was exactly along the galactic major (minor) axis, and for
this reason, we averaged two cuts at $6.8\degr$ and $8.1\degr$ from
the major (minor) axis. Observations by other authors were
sufficiently close to the major (minor) axis.

Velocity dispersions and rotation velocities along the major axis
are presented in Figs.~\ref{bootdmaj} and~\ref{bootvmaj},
respectively. Individual observations have been averaged; for
error-bars a bootstrap method was used, combining bootstrap errors
with observational errors. Outside the major and the minor axis, the
observations by \citet{mcelroy:83} were combined into four cuts at
$23\degr$, $38\degr$, $53\degr$ and $68\degr$ from the major axis
and they are shown by filled circles in Fig.~\ref{dispbulge}.

\subsection{Planetary nebulae observations}

Kinematical observations for PN have been published by
\citet{halliday:06,merrett:06}. \citet{halliday:06} used 50~nebulae for every
data point and \citet{merrett:06} used roughly 40~nebulae. For every
data point, they have found the averaged velocity and dispersion
near the galactic major axis.

In our model, we handle the kinematics of PN in a similar way to the
kinematics of the stellar component. As a result, the stellar observations give
us the kinematics in
inner parts of the galaxy and PN observations give the kinematics
mainly in outer parts. In overlapping regions both data are in
accordance.

\subsection{Observations of red giant branch (RGB) stars.}

\citet{reitzel:02} observed the kinematics of RGB stars along the minor
axis. At a distance of 19\,kpc from the centre, the calculated
velocity dispersion was found to be
$150^{+50}_{-30}$\,$\rmn{km\,s^{-1}}$. \citet{kalirai:06} measured 106
RGB stars at 12\,kpc along the minor axis and the derived dispersion
was 90\,$\rmn{km\,s^{-1}}$. The most recent observations by
\citet{gilbert:07} involved distances in the range of 9--30\,kpc along
the minor axis. They measured RGB stars in different fields and the
averaged dispersion in this region is 129\,$\rmn{km\,s^{-1}}$. The
results from different fields and errors are presented in
Fig.~\ref{dispmin2}.

\citet{chapman:06} made observations of RGB stars along the major axis
in the halo region (10--70\,kpc). They made spectroscopic
observations for 827 stars. The main conclusion was that the halo
sample shows no significant evidence for rotation. They measured the
central dispersion 152\,$\rmn{km\,s^{-1}}$, which decreases with
radius. The observed dispersions by \citet{chapman:06} are higher than
stellar or PN observations. There is no strict explanation for that,
but probably the line-of-sight velocity component is not correctly
subtracted, and therefore the dispersions are higher. Also, there is
a possibility that RGB stars have systematically higher dispersions.
Unfortunately, in our final model we cannot compare these
observations with our model output.

\section{A dynamical model}\label{sec:3}

\subsection{Calculation of stellar velocities and
dispersions}\label{sec:calcveldisp}

Details of the dynamical model described below can be found in
\citet{tempel:06}. However, to improve the readability of the paper, a
brief outline is given here.

\subsubsection{Basic formulae}

Let $(R, z, \theta )$ be cylindrical coordinates and
$a=\sqrt{R^2+z^2/q^2}$, where $q$ is the axial ratio of isodensity
ellipsoids. Knowing spatial luminosity densities of the components
$l_i(a)$ and ascribing a $M/L$-ratio to each component $f_i$
($i$~indexes the bulge,
the disc, the
stellar halo and the outer diffuse stellar halo; see Paper~I), we
obtain the spatial mass density distribution of a galaxy
\begin{equation}
\rho (a) = \sum_{i=1}^5 f_i l_i(a) + \rho_{\rmn{DM}}(a) \label{eq3}
\end{equation}
($\rho_{\rmn{DM}} (a)$ is the DM density). On the basis of
spatial mass density distributions, derivatives of the gravitational
potential $\frac{\upartial \Phi}{\upartial R}$ and $\frac{\upartial
\Phi}{\upartial z}$ can be calculated.

In stationary collisionless stellar systems with axial-symmetry the
Jeans equations in cylindrical coordinates can be written in a
convenient form for further calculations
\begin{equation}
\frac{\upartial \rho\sigma^2_R}{\upartial R} \!+\! \left(\!
\frac{1\!-\!k_\theta}{R} +
\frac{\upartial \kappa}{\upartial z} \!\right)\rho\sigma^2_R \!+\!
\kappa\frac{\upartial \rho\sigma^2_R}{\upartial z} \!=\! -\rho\frac{\upartial
\Phi}{\upartial R} \left(
\!1\!-\!\beta^2 \right)\!, \label{jj1}
\end{equation}
\begin{equation}
\frac{\upartial \rho\sigma^2_z}{\upartial z} + \left( \frac{\xi}{R} +
\frac{\upartial \xi}{\upartial R}
\right)\rho\sigma^2_z + \xi\frac{\upartial \rho\sigma^2_z}{\upartial R} =
-\rho\frac{\upartial \Phi}{\upartial z}, \label{jj2}
\end{equation}
where
\begin{equation}
\kappa\equiv\frac{1-k_z}{2}\tan(2\alpha), \qquad \xi\equiv\frac{\kappa}{k_z},
\end{equation}
\begin{equation}
k_z\equiv\frac{\sigma^2_z}{\sigma^2_R}, \qquad
k_\theta\equiv\frac{\sigma^2_\theta}{\sigma^2_R}.
\end{equation}
$\alpha$ is the angle between the major axis of the velocity
dispersion ellipsoid and the galactic plane. For each component, the
rotation velocity has been taken $V_{\theta} = \beta V_{\rmn c}$,
where $V_{\rmn c}$ is circular velocity and $\beta$ is a constant
specific for each subsystem.

The Jeans equations~(\ref{jj1})~and~(\ref{jj2}) contain unknown
functions $k_z$, $k_\theta$ and $\alpha$. In our model, the
phase-density of a system is a function of three integrals of
motion. Under this assumption, we can derive the unknown functions.

From the Jeans equations~(\ref{jj1})~and~(\ref{jj2}) the dispersions
along the coordinate axis ($\sigma_R$, $\sigma_z$ and
$\sigma_\theta$) can be calculated
\begin{equation}
\rho\sigma^2_R(R,z)=\!(1-\beta^2)\!\!\int\limits^\infty_R
\!\rho\frac{\upartial \Phi(r,z)}{\upartial R}\!\!\left[\exp\!\! \int\limits^r_R
p(r^\ast,z)\rmn{d}r^\ast\right]\!\!
\rmn{d}r,\\
\end{equation}
\begin{equation}
\rho\sigma^2_z(R,z)=\!\!\int\limits^\infty_z
\rho\frac{\upartial \Phi(R,z')}{\upartial z}\left[ \exp\!\! \int\limits^{z'}_z
g(R,z^\ast)\rmn{d}z^\ast \right]\rmn{d}z',
\end{equation}
\begin{equation}
\sigma^2_\theta=k_\theta\sigma^2_R \qquad \rmn{or} \qquad
\sigma^2_\theta=\frac{k_\theta}{k_z}\sigma^2_z,
\end{equation}
where
\begin{equation}
p\equiv\frac{1-k_\theta}{R}+\frac{\upartial \kappa}{\upartial z} ,
\qquad g\equiv\frac{\xi}{R}+\frac{\upartial \xi}{\upartial R}.
\end{equation}

The details of derivation and some restrictions can be found in
\citet{tempel:06}.

\subsubsection{Line-of-sight dispersions}

\begin{figure}
\includegraphics[width=84mm]{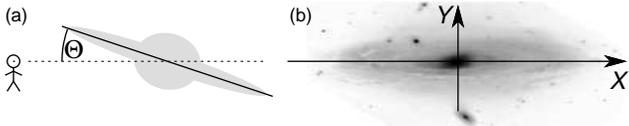}
\caption{Panel~(a) gives the angle between the line-of-sight and the
galactic plane. Panel~(b) shows our galactic coordinates $X$ and $Y$
in the sky.} \label{figlosgal}
\end{figure}

Calculated velocity dispersions $\sigma^2_R$, $\sigma^2_z$ and
$\sigma^2_\theta$ cannot be directly compared with observations. The
calculated dispersions must be projected to the line-of-sight.
Designating $\Theta$ as the angle between the line-of-sight and the
galactic plane (see Fig.~\ref{figlosgal}), the line-of-sight
dispersions $\sigma^2_{\rmn{los}}$ are
\begin{equation}
\sigma^2_{\rmn{los}}=\sigma^2_\ast\cos^2\Theta + \sigma^2_z\sin^2\Theta ,
\end{equation}
where
\begin{equation}
\sigma^2_\ast=\sigma^2_\theta\frac{X^2}{R^2} + \sigma^2_R \left(
1-\frac{X^2}{R^2}\right) .
\end{equation}

The next step is to integrate individual line-of-sight velocity
components along the whole line-of-sight
\begin{equation} \label{final_eq}
\sigma^2_{\rmn{int}}(X,Y)=\frac{1}{L(X,Y)}\int\limits^{\infty}_{X}
R\frac{\sum\limits_{i=1}^2\left[l(R,z_i)\sigma^2_{\rmn{los}}(R,z_i)\right]}{\cos
{
\Theta}
\sqrt{R^2-X^2}}\rmn{d}R,
\end{equation}
\begin{equation}
z_{1,2}= \left( \frac{Y}{\sin{\Theta}} \pm \sqrt{R^2-X^2} \right)
\tan{\Theta},
\end{equation}
where $l(R,z)$ denotes galactic spatial luminosity density, and
$L(X,Y)$ is the surface luminosity density profile.

Equation~(\ref{final_eq}) gives the line-of-sight dispersion for one
galactic component. Our model consists of several components and we
must sum over all components considering the surface luminosity
profile
\begin{equation}\label{fin_disp}
\sigma^2_{\rmn{obs}}(X,Y)=\frac{\sum\limits_i\left\{L_i(X,Y)
\left[\sigma^2_{\rmn{int}}(X,Y)\right]_i\right\}}{\sum\limits_i L_i(X,Y)},
\end{equation}
where $i$ denotes the subsystem and the summation is taken over all
subsystems.

\subsubsection{Line-of-sight stellar velocities}

In an axisymmetric system, we have only one velocity component, the
rotational velocity $V_\theta$. According to the definition of the
circular velocity, we can write for each component
\citep[][equation~(2.19)]{binney:87}
\begin{equation}
V_\theta^2=\beta^2R\frac{\upartial \Phi}{\upartial R} = \beta^2 4\upi
Gq\int\limits_0^R
\frac{\rho (a) x^2\rmn{d}x}{\sqrt{R^2-e^2x^2}},
\end{equation}
where $a^2 = x^2+z^2x^2(R^2-e^2x^2)^{-1}$ and $e=\sqrt{1-q^2}$ is
eccentricity. The rotational velocity cannot be compared directly to
the observed velocity. We use the same projection method for stellar
velocities as we used for dispersions. The projected line-of-sight
velocity is
\begin{equation}
V_{\rmn{los}}=V_\theta\frac{X}{R}\cos{\Theta}.
\end{equation}
The observable velocity $V_\rmn{obs}$ is
\begin{equation}\label{fin_rot}
V_{\rmn{obs}}(X,Y)=\frac{\sum\limits_i\left\{L_i(X,Y)
\left[V_\rmn{int}(X,Y)\right]_i\right\}}{\sum\limits_i L_i(X,Y)},
\end{equation}
where $V_\rmn{int}$ is the same as the right-hand side of
equation~(\ref{final_eq}), only $\sigma^2_\rmn{los}\equiv
V_\rmn{los}$. The summation is taken over all subsystems.

\subsection{Calculation of dark matter distribution}\label{sec3p2}

Total gravitational potential is a sum of the potentials caused by
visible matter and DM. In terms of circular velocities,
\begin{equation} \label{Vcobs}
V_\rmn{c}^{^2\!\rmn{tot}}=V_\rmn{c}^{^2\!\rmn{vis}}+V_\rmn{DM}^{^2} .
\end{equation}

Photometrical and chemical evolution models give us the spatial mass
density of visible matter (see equation~(\ref{eq3})) and thereafter
$V_\rmn{c}^{^2\!\rmn{vis}}$. If gas velocity dispersions are small in
comparison with rotation velocities and we neglect also non-circular
motions, gas rotation velocities can be identified with total
circular velocities and equation~(\ref{Vcobs}) gives us the spatial
mass density distribution of DM
\begin{equation} \label{darkmatterrho}
\rho_\rmn{DM}(R)=\frac{1}{4\upi
R^2G}\frac{\rmn{d}}{\rmn{d}R}\left(RV_\rmn{DM}^{^2}\right)
\end{equation}
and inner mass of DM
\begin{equation} \label{darkmattermass}
M_\rmn{DM}(R)=\frac{RV_\rmn{DM}^{^2}}{G},
\end{equation}
where $G$ is the gravitational constant.

Unfortunately, this straightforward calculation is too simplified.
In the most interesting bulge region, ignoring of non-circular
motions and dispersions of the gas is not justified. For this
reason, total circular velocities remain unknown and we must use
another way to calculate the DM density.

In a simple form, the first Jeans equation~(\ref{jj1}) can be
written as
\begin{equation} \label{jeansgeneral}
V_\rmn{c}^2=V_\theta^2+\sigma_R^2 f(R) ,
\end{equation}
where $f(R)$ is a specific function. Applying~(\ref{jeansgeneral})
for total matter (visible\,+\,DM), we have
\begin{equation}
V_\rmn{c}^{^2\!\rmn{tot}}=V_\theta^{^2\!\rmn{tot}}+ \sigma_R^{^2\!\rmn{tot}}
f(R) .
\end{equation}
Substituting the last equation into equation~(\ref{Vcobs}), we get
\begin{equation} \label{Vcdmstar}
V_\rmn{DM}^{^2}=V_\theta^{^2\!\rmn{tot}}-V_\rmn{c}^{^2\!\rmn{vis}}+
\sigma_R^{^2\!\rmn{tot}}
f(R) ,
\end{equation}
where $\sigma_R^{^2\!\rmn{tot}}$ and $V_\theta^\rmn{tot}$ are the observed
velocity dispersion and the observed rotational velocity,
respectively. The circular velocity of the visible matter results
from the component parameters derived in Paper~I. It is reasonable
to assume that the function $f(R)$ is the same for total matter and
for visible matter (DM is collision-free). Taking this into account,
\begin{equation}
V_\rmn{c}^{^2\!\rmn{vis}}=V_\theta^{^2\!\rmn{vis}}+ \sigma_R^{^2\!\rmn{vis}}
f(R) .
\end{equation}
Expressing from the previous equation $f(R)$, we can write
\begin{equation} \label{fReq}
f(R)=\frac{V_\rmn{c}^{^2\!\rmn{vis}}-V_\theta^{^2\!\rmn{vis}}}
{\sigma_R^{^2\!\rmn{vis}}} ,
\end{equation}
where $V_\theta^\rmn{vis}$ and $\sigma_R^{^2\!\rmn{vis}}$ are the visible
matter rotational velocity and dispersion, respectively. We can find
these quantities, using the method described in
Section~\ref{sec:calcveldisp}. Substituting equation~(\ref{fReq})
into equation~(\ref{Vcdmstar}), we finally get the DM circular
velocity
\begin{equation} \label{Vcdmfinal}
V_\rmn{DM}^{^2}=V_\theta^{^2\!\rmn{tot}}-V_\rmn{c}^{^2\!\rmn{vis}}+
\frac{\sigma_R^{^2\!\rmn{tot}}}{\sigma_R^{^2\!\rmn{vis}}}
\left(V_\rmn{c}^{^2\!\rmn{vis}}-V_\theta^{^2\!\rmn{vis}}\right) .
\end{equation}
Using equations (\ref{darkmatterrho}), (\ref{darkmattermass}) and
(\ref{Vcdmfinal}), we find the DM density distribution and the inner
mass of DM, taking into account the rotation and dispersion data.

Rotational velocities and dispersions in equation~(\ref{Vcdmfinal})
must be determined in the galactic equatorial plane as a function of
radius.

The Jeans equation in form~(\ref{jeansgeneral}) is not entirely
correct, because we did not consider dispersion derivatives. To have
the correct form, an additional term must be added to the right-hand
side
\begin{equation} \label{jeanslisaliige1}
-R\left[\frac{\sigma^2}{\rho}\frac{\upartial \rho}{\upartial R}+\frac{\upartial
\sigma^2}{\upartial R}\right].
\end{equation}
Taking this correction into account, also a correction to the
right-hand side of equation~(\ref{Vcdmfinal}) must be added
\begin{equation} \label{jeanslisaliige2}
R\!\left[\!\frac{\sigma^2_\rmn{obs}}{\sigma^2_\rmn{vis}}\frac{\upartial
\sigma^2_\rmn{vis}}{\upartial R}
\!-\! \frac{\upartial \sigma^2_\rmn{obs}}{\upartial R}\!\right]\! +
R\sigma^2_\rmn{obs}\!\left[\!\frac{1}{\rho_\rmn{vis}}\frac{\upartial
\rho_\rmn{vis}}{\upartial R} \!-\!
\frac{1}{\rho_\rmn{tot}}\frac{\upartial \rho_\rmn{tot}}{\upartial R}\!\right]\!.
\end{equation}
This additional term is important only on a small scale. We cannot
adequately consider the derivatives of dispersions as numerical
differentiation amplifies noises in observational data too much, and
therefore we do not use this term in our calculation.

\section{Mass distribution model of M\,31}\label{sec:4}

We assume M\,31 to consist of the following subsystems: a bulge,
a disc, a
stellar halo, an outer diffuse stellar halo and a DM halo. Reasoning of
the component selection (except DM halo) and their parameters
resulting from the photometrical model
are given in table~1 in Paper~I. Luminosity distributions of the
components are transformed to mass distributions on the basis of
$M/L$'s resulting from chemical evolution models.
Chemical evolution
models give the components most probable $M/L$ in different colours
and possible $M/L$ ranges.

As most kinematical observations have been made at wavelengths close
to the $V$-band, the $V$-luminosities are used in weighting the
components if superposition is needed
(Equations (\ref{fin_disp}) and (\ref{fin_rot})). Component
parameters most relevant in context of the present paper are given
in Table~\ref{model_param2}. According to \citet{nieten:06}, we add to
the disc a total neutral gas mass
$5.2\cdot10^9\rmn{M_{\sun}}$.

Dynamical models can further constrain the permitted mass intervals.
In our dynamical modelling process, we tried many different
models, and, in most cases, the main conflict was that chemical
evolution models predict larger bulge masses than dynamical
models permit. The masses of other components remain between
the permitted values. In our final dynamical model, masses
for the disc, the stellar halo and the outer diffuse stellar halo
are in accordance with the masses from the chemical evolution model.
To achieve the correct concentration parameter for the NFW
dark matter halo, we slightly decreased the best disc mass resulting
from the chemical evolution models. The bulge mass in our final
model is also slightly smaller than the minimal stellar mass
predicted by the chemical evolution model. The smaller bulge mass
was needed for an agreement with the observed velocity dispersion
data in the bulge region.

\begin{table}
\caption{Calculated model parameters for stellar components.}
\label{model_param2}
\begin{tabular}{@{}llllllllllll}
\hline
Popul.   & $L_V$& $a_0$ & $q$  & $N$  &$M/L_V$& $M$ &$\beta$ \\
\hline
Bulge    & 1.45 & 0.64  & 0.6  & 4.2 &4.2$_{-1.4}^{+1.0}$   &3.39 & 0.25\\
Disc     & 2.02 & 9.3   & 0.05 & 0.7 &3.1$_{-0.7}^{+1.2}$   &5.57 & 0.98\\
Halo     & 0.61 & 4.0   & 0.5  & 2.7 &2.9$_{-0.7}^{+1.7}$   &1.76 & 0.4\\
Dif. halo& 0.05 & 40.0  & 0.9  & 2.0 &2.1$_{-0.6}^{+0.6}$   &0.11 & 0.3\\
\hline
\end{tabular}

\medskip
Masses and luminosities are in units of $10^{10}\rmn{M_{\sun}}$ and
$10^{10}\rmn{L_{\sun}}$ respectively; component radii are in kpc; $M/L_V$ is
expected $M/L$-ratio from chemical evolution models and $M$ is mass used in our
dynamical model.
\end{table}

In addition to masses, dynamical models contain a parameter $\beta$.
This parameter was, in fact, determined for each component on the
basis of fitting the stellar and PN rotation velocities with the
model. On the other hand, an increasing of rotation velocities
causes a decrease of velocity dispersions. In this way also
dispersions influence the determination of $\beta$. Final values of
$\beta$ parameters are given in Table~\ref{model_param2}.

\begin{figure}
\includegraphics[width=84mm]{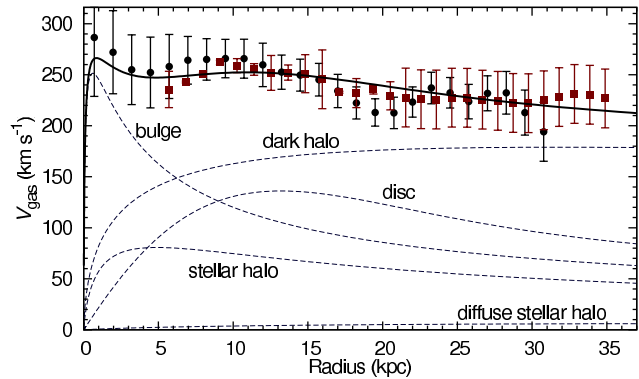}
\caption{Gas rotational velocity. The filled circles are an averaged
profile (see Fig.~\ref{bootgasvel}), the filled squares show the
rotation curve from \citet{carignan:06}. The solid line denotes circular
velocity derived from our model. The dashed lines show the
rotational velocity for different components.}\label{gasvelo}
\end{figure}
\begin{figure}
\includegraphics[width=84mm]{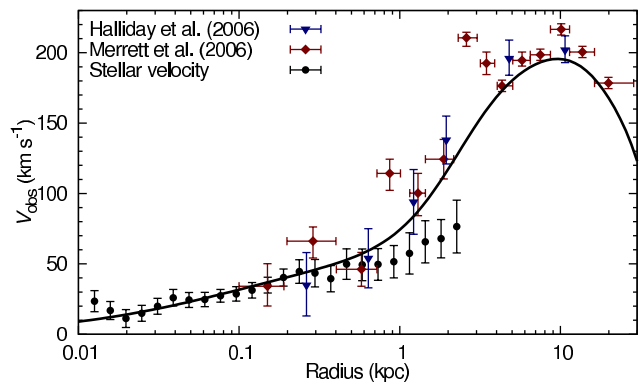}
\caption{PN and stellar velocity along the major axis. The filled
triangles and the diamonds are PN rotational velocities, the filled
circles are averaged stellar velocities. The solid line is our
calculated stellar velocity.}\label{velo}
\end{figure}
\begin{figure}
\includegraphics[width=84mm]{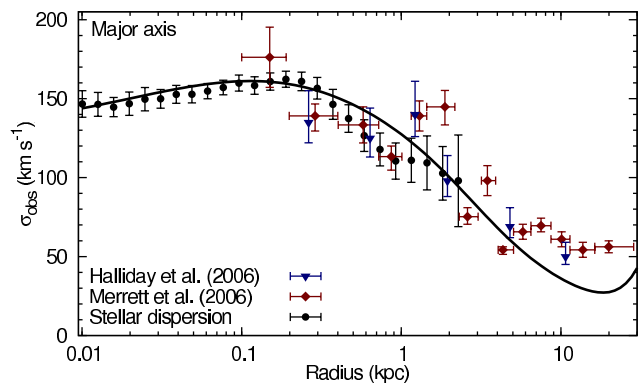}
\caption{Dispersions along the major axis. The filled triangles and
the diamonds are PN dispersion observations. The filled circles are
averaged stellar dispersions. The solid line is the dispersion from
our model.}\label{dispmajor}
\end{figure}
\begin{figure}
\includegraphics[width=84mm]{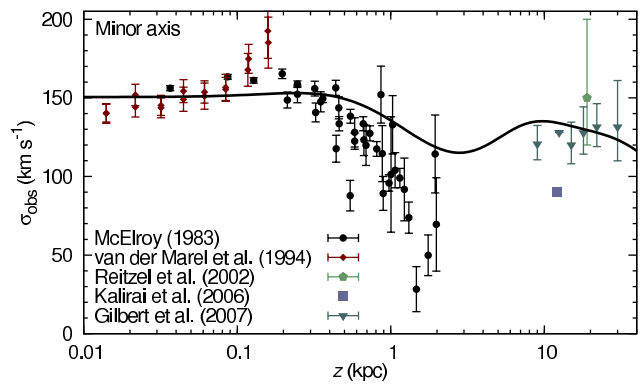}
\caption{Stellar dispersions along the minor axis. The filled
circles and the diamonds are stellar dispersions. The pentagon,
the square and the triangles are individual RGB stars observations. The
solid line is our modelled dispersion.}\label{dispmin2}
\end{figure}
\begin{figure}
\includegraphics[width=84mm]{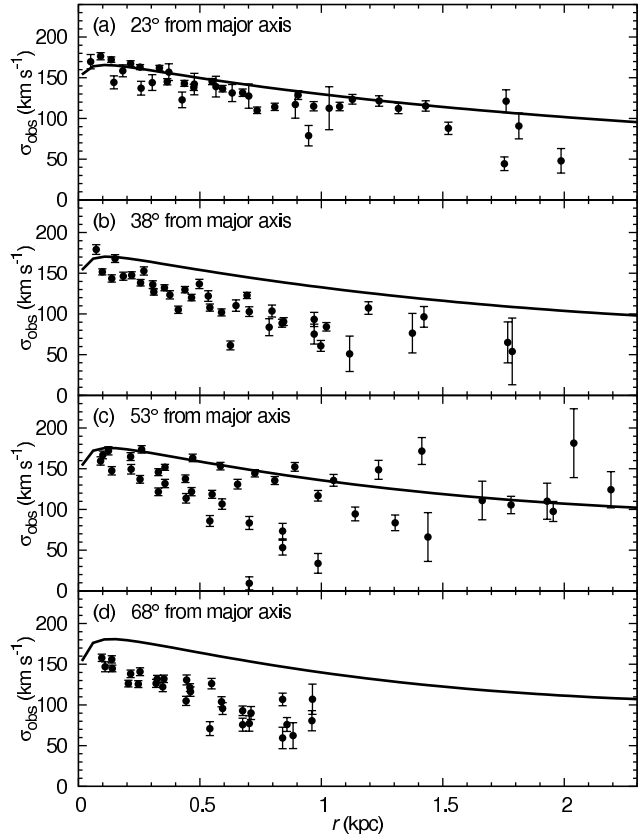}
\caption{Dispersions along slits between the major and the minor
axis. The filled circles are dispersions from \citet{mcelroy:83}. The solid
lines are our modelled dispersions. The distance $r$ is along the
slit.}\label{dispbulge}
\end{figure}

Circular velocities for visible and DM components in the final model
are given in Fig.~\ref{gasvelo} as dashed lines. For comparison,
observed gas rotation velocities are given by filled circles and
filled squares. Our model is in good agreement with observations.

Stellar rotational velocity is given in Fig.~\ref{velo}. In the
inner region ($R<0.02$\,kpc), the observational velocities are
higher than the calculated velocities. M\,31 has a fast rotating
nucleus and the increase of velocities can be explained by using the
nucleus. Our model does not include the nucleus and therefore we
have a lower velocity profile in this region. The averaged observed
stellar rotational velocities in the outer
region ($R>0.7$\,kpc) are
smaller than the calculated velocities. We do not have a good
explanation for that. However, the observations are very uncertain
in this region (see Fig.~\ref{bootvmaj}) and some observations are
in agreement with our calculated results. We have a satisfactory fit
with PN observations. A small disagreement is only in the outer regions,
where the calculated velocities are slightly smaller than the observed ones.

Calculated velocity dispersions along the major axis are given in
Fig.~\ref{dispmajor}. It is seen that we have a good fit with
observations. Only in the outer region ($R>5$\,kpc, where
only PN observations are available) the calculated dispersions are
slightly smaller than the observed dispersions. The probable reason
for that is that PN observations do not lie exactly along the
galactic major axis. The second reason is that the observed
dispersions include a component caused by rotation at different
positions along the line of sight, which increases the observed
dispersions by up to twenty kilometers per second.

Calculated velocity dispersions along the minor axis are presented
in Fig.~\ref{dispmin2}. In addition to usual stellar dispersion
measurements, we have plotted in Fig.~\ref{dispmin2} also
independent  dispersion measurements for RGB stars by
\citet{reitzel:02,kalirai:06,gilbert:07}, extending to larger distances $z$ from
the galactic plane. The dispersions calculated from our best-fitting
model are in good agreement with observations of RGB stars. In the
intermediate regions, the calculated dispersions lie higher than the
observed ones. It is possible to have a better fit by using the
elliptical DM halo, but still the calculated dispersions are greater
than the observed ones.

In Fig.~\ref{dispbulge} the observed and calculated dispersions
along the lines between the major and the minor axis (at $23\degr$,
$38\degr$, $53\degr$ and $68\degr$ from major axis) are presented.
The observations are with high scatter and they are not symmetric
with respect to the galactic centre and are thus uncertain.

\section{Dark matter distribution in M\,31}\label{sect_dm}

One of the aims of Paper~I and the present paper is to derive
constraints to the possible DM density distribution and to compare
these constraints with several known analytical spherical DM density
profiles.

First, we chose for comparison a typical cored density
distribution \citep{burkert:95}
\begin{equation}
\rho_\rmn{Burkert}(r)=\frac{\rho_0}{\left(1+\frac{r}{r_\rmn{c}}\right)
\left[1+(\frac{r}{r_\rmn{c}})^2\right]} ,
\end{equation}
where $\rho_0$ is central density and $r_\rmn{c}$ is characteristic
radius.
The cored isothermal density distribution profile is rather
similar to the Burkert profile (with the exception of the outer
parts) and we left it aside. Next we also used different cuspy
density distributions: the Moore profile \citep{moore:99}
\begin{equation}
\rho_\rmn{Moore}(r)=\frac{\rho_\rmn{c}}{(\frac{r}{r_\rmn{c}})^{1.5}
\left[1+(\frac{r}{r_\rmn{c}})^{1.5}\right]} ,
\end{equation}
and the NFW profile \citep{navarro:97}
\begin{equation}
\rho_\rmn{NFW}(r)=\frac{\rho_\rmn{c}}{\left(\frac{r}{r_\rmn{c}}\right)
\left[1+(\frac{r}{r_\rmn{c}})\right]^{2}} .
\end{equation}
In the Moore and NFW profiles, $\rho_\rmn{c}$ is a density scale
parameter.

 Later \citet{navarro:04} have found instead of the NFW profile
another density distribution law which fits over a wider distance
range. According to \citet{merritt:06}, we refer to this equation
as ``Einasto's $r^{1/n}$ model'' \citep{einasto:65,einasto:68,einasto:69}.
Following \citet{merritt:06}, we replace the exponent $\alpha$ in
\citet{navarro:04} equation by $1/n$
\begin{equation}\label{EinastoModel}
\rho_\rmn{Einasto}= \rho_\rmn{c}\exp\left\{-d_{n}
\left[\left(\frac{r}{r_\rmn{c}}\right)^{1/n}-1\right]\right\},
\end{equation}
where $n$ is in principle a free parameter. According to N-body
simulations, we can take $n = 6.0\pm 1.1$ \citep{navarro:04}. The term
$d_n$ is a function of $n$ in a way that $\rho_{\rmn{c}}$ is the
density at $r_{\rmn{c}}$ defining a volume containing half of the
total mass. The value of $d_n$ can be well approximated by the
expression \citep{merritt:06}
\begin{equation}
d_n  \approx 3n-1/3+0.0079/n.
\end{equation}

 The NFW density profile has extensively been studied and a
useful characteristic for this profile is the concentration
parameter, $c_{\rmn{vir}}$, defined as the ratio between the virial
and inner radii,
\begin{equation}
 c_{\rmn{vir}}\equiv R_{\rmn{vir}} / r_{\rmn{c}}.
\end{equation}
The outer, virial radius $R_{\rmn{vir}}$ of a halo of the virial
mass $M_{\rmn{vir}}$, is defined as the radius within which the mean
density is $\Delta_{\rmn{vir}}$ times the critical density of the
universe $(\rho_{\rmn{crit}}=3H_0^2/8\pi G)$:
\begin{equation}
 M_{\rmn{vir}} \equiv
\frac{4\pi}{3}\Delta_{\rmn{vir}}\rho_{\rmn{crit}}R_{\rmn{vir}}^3.
\end{equation}
In the $\Lambda\rmn{CDM}$ cosmological model, the local value is
$\Delta_{\rmn{vir}}\simeq 337$.

 Using N-body simulations, the correlation between the halo
virial mass and the concentration parameter for the
$\Lambda\rmn{CDM}$ model has been found by
\citet{bullock:01,wechsler:02}. For an Andromeda-sized halo, the
concentration parameter is in the range $10<c_{\rmn{vir}}<20$. In
our model, the concentration parameter is 11.7, which is in
accordance with N-body simulations. In principle, in the case of
Moore's law, the concentration index should also  be considered.
Unfortunately, the dependence of the concentration index of Moore's
law on DM mass is unknown and we cannot take it into account. For
the Einasto law, the consentration index in our model is similiar to
that for NFW law.

Following the scheme presented in Section~\ref{sec3p2}, we can
estimate the allowed region for the DM density distribution in
M\,31.

\begin{figure}
\includegraphics[width=84mm]{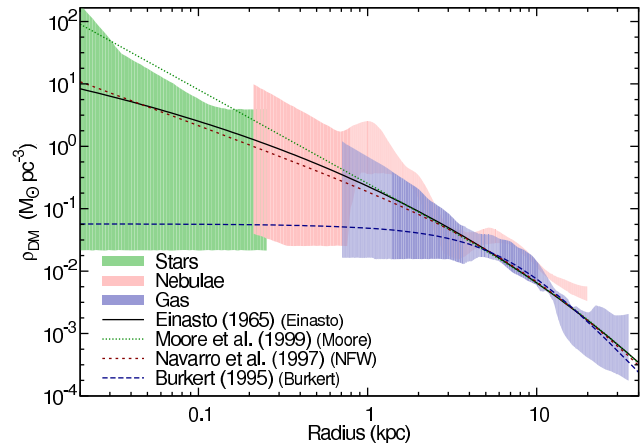}
\caption{DM density profile. The filled areas are density profiles
derived form stars, PN and gas data. Different lines represent
different DM profiles. In our final model we have used the Einasto
profile.}\label{dm density}
\end{figure}

In Fig.~\ref{dm density} the derived DM density distribution as a
function of radius is given. To achieve the maximum range in radius,
we used three different types of data: stellar observations, PN
observations and the gas rotation curve. From kinematical data
(Fig.~\ref{velo} and~\ref{dispmajor}) we see that PN give us the
mass distribution in an intermediate range 0.2--20\,kpc. Using
stellar observations, we can obtain the mass distribution in inner region. In
the outer region (up to 35\,kpc), we estimate the mass
distribution using the gas rotation curve. Using three different
types of observations we obtain the DM mass distribution in the
range 0.02--35\,kpc. The mass distributions derived from PN data
have an intersection with stellar and gas data. In interlapping
regions all three estimates are in accordance. The error-bars
(filled areas) in Figs.~\ref{dm density} and~\ref{dm density int}
are derived using kinematical observational errors only. The real
error-bars are slightly larger, because we must add errors from
photometrical observations and intrinsic errors of the model.

Derived in this way density distribution has been fitted with the
analytical profiles referred above. Best-fitting profiles are
plotted also in Fig.~\ref{dm density} and~\ref{dm density int}, the
corresponding parameters of the profiles are given in
Table~\ref{dm_dat}. In addition, virial masses
$M_{\rmn{vir}}$ and virial radii $R_{\rmn{vir}}$ for each profile
are given. In our final model (described in the previous section),
we have used the Einasto profile.

\begin{table}
\caption{Dark matter profiles.} \label{dm_dat}
\begin{tabular}{@{}lllll}
\hline
Profile      & $\rho_0$, $\rho_\rmn{c}$  & $r_\rmn{c}$ & $M_{\rmn{vir}}$ &
$R_{\rmn{vir}}$\\
             & $[\rmn{M_{\sun}}\rmn{pc}^{-3}]$ & [kpc] &
$[10^{11}\rmn{M_{\sun}}]$ & [kpc]\\
\hline
Einasto$^{a}$& $1.35\cdot 10^{-5}$ & 135.0& 7.85 & 152.7\\
NFW          & $1.74\cdot 10^{-2}$ & 12.5 & 6.93 & 146.5\\
Moore        & $2.05\cdot 10^{-3}$ & 25.0 & 7.38 & 149.6\\
Burkert      & $5.72\cdot 10^{-2}$ & 6.86 & 5.16 & 132.8\\
\hline
\end{tabular}
\begin{description}{}{}
\item[$^a$] Parameter $n$ has been taken 5.8.
\end{description}
\end{table}

On the basis of Fig.~\ref{dm density}, the inner mass of the DM
component was calculated and plotted in Fig.~\ref{dm density int} as
a function of radius. A number of researchers have used different
objects to estimate the total mass inside a given radius or the
total mass of M\,31. In Table~\ref{table_totmass} we present some
mass estimates at different radii with error-bars. According to
\citet{geehan:06}, most realiable estimates are from \citet{evans:00b},
using the PN and the GC data and an estimate by \citet{geehan:06} on
the basis of satellite galaxies. At a distance
$r_\rmn{max}=125$\,kpc three independent mass estimates have been
made with the mean value $7.5\cdot 10^{11}\rmn{M_{\sun}}$. 
The referred mass estimates in Fig.~\ref{dm density int} correspond
to dark matter masses: from total masses given in
Table~\ref{table_totmass} the stellar mass was subtracted.
\begin{table}
\caption{Mass estimates for M\,31 (out to a radius $r_\rmn{max}$).}
\label{table_totmass}
\begin{tabular}{@{}llll}
\hline
Reference               & $r_\rmn{max}$ & Mass              & Objects \\
                        & $\rmn{[kpc]}$& $[10^{10}\rmn{M_{\sun}}]$&    \\
\hline
\citet{perrett:02}         & 27     & $41^{+1}_{-1}$      & GC \\
\citet{evans:00b}$^{\ast}$   & 31     & $28^{+24}_{-12}$    & PN \\
\citet{carignan:06}$^\ast$   & 35     & $34^{}_{}$          & H\,{\sc i} \\
\citet{evans:00b}$^{\ast}$   & 40     & $47^{+34}_{-23}$    & GC \\
\citet{lee:08}           & 55     & $55^{+4}_{-3}$      & GC \\
\citet{galleti:06}$^\ast$& 60     & $44^{+2}_{-2}$      & GC \\
\citet{ibata:04}$^\ast$  & 125    & $75^{+25}_{-13}$    & stream \\
\citet{chapman:06}         & 125    & $72^{}_{}$          & RGB stars\\
\citet{fardal:06}$^\ast$  & 125    & $74^{+12}_{-12}$    & stream \\
\citet{geehan:06}$^\ast$  & 158    & $55^{+71}_{-31}$    & satellites \\
\citet{evans:00a}         & 500    & $70^{+105}_{-35}$   & satellites \\
\citet{evans:00b}            & total  & $123^{+180}_{-60}$  & satellites \\
\hline
\end{tabular}
\begin{description}{}{}
\item[$^*$] These points are shown in Fig.~\ref{dm density int}.
\end{description}
\end{table}
\begin{figure}
\includegraphics[width=84mm]{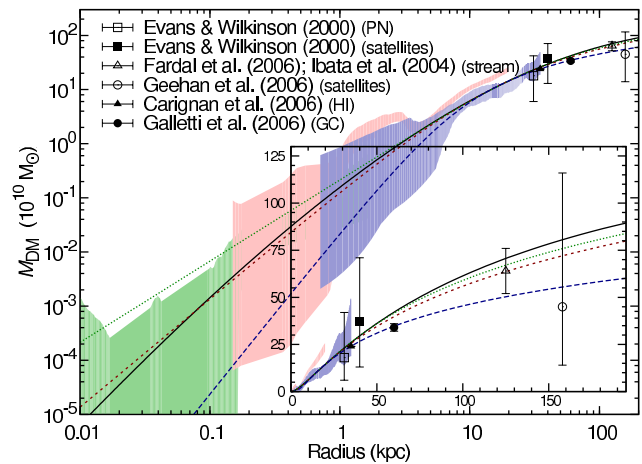}
\caption{Integrated DM density profile. The filled areas and the
lines are the same as in Fig.~\ref{dm density}. The points are mass
estimates from different authors using different objects.}\label{dm
density int}
\end{figure}

When comparing the derived DM distribution with most common
analytical DM density distributions (Figs.~\ref{dm density}
and \ref{dm density int}), it is seen that within uncertainties
nearly all distributions fit with the limits from observations. On
the other hand, it is also seen that the Moore profile lies in
central regions slightly outside the upper limits of the
uncertainties; the Burkert profiles lie rather close to the
lower limits of the uncertainties. The Burkert formula
predicts also a slightly smaller total and virial mass than
the cuspy profiles. It is seen that the NFW and the Einasto
density distributions give the best fit.

\section{Discussion}\label{sec:6}

In Figs.~\ref{dm density} and~\ref{dm density int} we tried to
estimate all possible uncertainties. The contribution of errors in
photometrical surface brightness profiles is small. There is some
uncertainty in component subtraction but its influence due to
integration along the line-of-sight is relatively small. The most
significant uncertainty arises from the stage of using chemical
evolution models. First, different colour indices are not in good
mutual agreement. Second, chemical evolution models involve several
free parameters (initial mass function (IMF), age, star formation
duration, contribution of stellar winds etc.). However, it was
possible to decrease the influence of all these uncertainties in our
final results by using several colour indices $(U-B)$, $(B-V)$,
$(V-R)$, $(R-I)$ and $(I-L)$. In addition, we used independent
metallicity measurements allowing us to significantly narrow the possible
models. The uncertainties resulting from chemical evolution models
were estimated by constructing a large amount of different models
with different initial conditions. All this allows us to ascribe
errors to the final $M/L$ and thereafter to the $M$ values of
visible matter for all components (see Table~\ref{model_param2}).

We derived that within uncertainties nearly all the commonly used
analytical DM density profiles fit with limits from observations
at nearly all distances. The best-fitting models, however,
are the NFW and the Einasto density profiles. Usually in
similar studies it is concluded that the NFW profile does not fit
with observations.

A common approach in this kind of studies at present is to select
galaxies without the bulge component or with a small bulge
\citep*[see e.g.][]{simon:05,kassin:06,gentile:04,gentile:07}. The gas rotation
curve is corrected for non-circular motions and expansion as well as
possible. Colour indices of the disc are derived from surface
brightness distributions. Using a colour-$M/L$ relation (usually by
\citet{bell:01}, updates \citet{bell:03a}) the mass distribution of
visible matter is derived. Identifying gas rotation velocities with
circular velocities it is possible to calculate the DM density
distribution.

To derive the $M/L$ of visible matter, quite frequently only one or
two colour indices are used. This may involve significant
uncertainty in the density distribution of visible matter and
thereafter in DM.

\begin{figure}
\includegraphics[width=84mm]{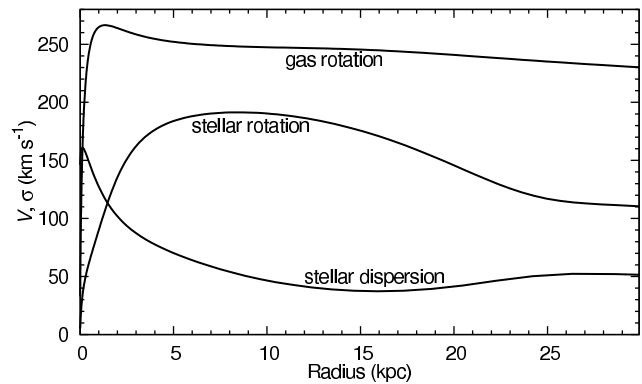}
\caption{Gas rotation, stellar rotation and stellar dispersion
components of a typical spiral galaxy as a function of the galactic
radius.} \label{gasdispvel}
\end{figure}

When modelling S0-Sa-Sbc disc galaxies with a significant bulge
component, it is necessary to take into account a detailed
luminosity distribution of the bulge and the disc, finite
ellipticities of these components, and both the rotation and the
dispersion components in stellar kinematics. When ignoring stellar
velocity dispersions, the calculated gravitational potential is
smaller. Dispersion components in kinematics of a typical spiral
galaxy are more dominating in the central parts of the galaxies (see
Fig.~\ref{gasdispvel}), and thus in the central parts the
gravitational potential is underestimated and the resulting DM
density profile is flatter near the centre.

Even in a sample of low mass disc, the power-law index in the
central regions of galaxies varies highly \citep{simon:05}. Thus it
is worth not to limit DM density studies with bulgeless low surface
brightness (LSB) galaxies but to model also galaxies of intermediate
and high luminosities and of different morphological types. In LSB
and late morphological type galaxies star formation is delayed. This
may be caused by insufficiently deep potential well of incompletely
formed DM haloes. Thus, it is not surprising that DM density
profiles in these galaxies are dominantly not cuspy.
In addition, recent study by \citet{pizzella:08} rises the question
about the reliability of the use LSB galaxies as a tracer of mass distribution,
in particular in the central regions of LSB galaxies.

Modelled dispersions along the minor axis (Fig.~\ref{dispmin2}) and
between the major and the minor axis (Fig.~\ref{dispbulge}) do not
have a very good fit with the observed stellar dispersions.
Unfortunately, this bulge region has been observed only by
\citep{mcelroy:83} and there is a possibility that these observations
include some systematic errors that have not been taken into
account. On the other hand, our models have several free parameters.
In our final model, we use spherical DM density distributions. We
have tried elliptical DM profiles and in some aspects the elliptical
DM profile gives slightly better results. In our model we have also
parameter $\beta$, which determines the rotational velocity for
stellar components. In our model, $\beta$ is a constant for each
subsystem, but principally $\beta$ might be a function of $R$ and
$z$. As we do not have good quality stellar rotation curves outside
the galactic major axis, we do not know how $\beta$ varies. The last
but not least uncertainty in our model is the orientation of the
triaxial velocity ellipsoid. In our model, the velocity ellipsoid is
radially elongated and lies under the angles $\leq 30\degr$ with
respect to the galactic equatorial plane \citep{tempel:06}. The velocity
ellipsoid orientation is quite uncertain and the orientation can
affect the modelled dispersions quite significantly. In one
of our following papers we analyse this kind of dynamical models in
greater detail.

Our derived total mass inside 100\,kpc ($5.6\cdot
10^{11}\rmn{M_{\sun}}$ for NFW and $6.2\cdot 10^{11}\rmn{M_{\sun}}$
for the Einasto model) is comparable with mass estimates
from \citet*{klypin:02} ($8.5\cdot 10^{11}\rmn{M_{\sun}}$),
\citet{geehan:06} ($6.0\cdot 10^{11}\rmn{M_{\sun}}$) and \citet*{seigar:06}
($6.5\cdot 10^{11}\rmn{M_{\sun}}$). The ratio of visible matter to
total matter (inside 100\,kpc) in our model (0.10 for the
NFW and 0.092 for the Einasto model) is in good agreement with the
results from \citet{klypin:02} (0.105). On the contrary, the ratios from
\citet{geehan:06} (0.173) and \citet{seigar:06} (0.152) are somewhat
higher than ours: the main difference is that they have almost twice
as massive a disc as in our model. \citet{geehan:06} admit that they
have degeneracy between the disc and dark halo components: reducing
the disc mass and increasing the dark halo mass we get better fit
with our model. The ratio of the DM mass to the
visible mass inside the Holmberg radius ($\sim25\,\rmn{kpc}$ for
M\,31) is 1.75 for the NFW and the Einasto distributions.

The total mass of M\,31 with the NFW DM distribution
is $1.19\cdot 10^{12}\rmn{M_{\sun}}$, the ratio of the DM
mass to the visible mass is 10.0. For the Einasto DM distribution
these values are $1.28\cdot 10^{12}\rmn{M_{\sun}}$ and 10.8. Total
masses are calculated inside 400\,kpc, which is half way from M\,31
to the Milky Way. The total mass calculated from our model is in
good agreement with the total mass calculated by \citet{evans:00b} on the
basis of M\,31 satellites.

On the basis of the proper motions of the Local Group galaxies IC\,10 and M\,33
and assuming that these galaxies are bound to M\,31, \citet{brunthaler:07}
calculated a lower limit for the mass of M\,31 of $7.5\cdot
10^{11}\rmn{M_{\sun}}$. This mass estimate is in accordance with our mass
estimate based on cuspy DM profiles. The cored Burkert profile in our model
gives slightly smaller mass than this lower limit.

In central regions the density distribution of the DM can be approximated as
$\rho\sim r^{-\beta}$. From Fig.~\ref{dm density} we can conclude that
$0 \leqslant \beta \leqslant 1.5$. On the basis of a solution of the Jeans
equations \citet{hansen:04} derived that $1\leqslant\beta\leqslant 3$. Resulting
common region is rahter narrow $1\leqslant\beta\leqslant 1.5$.

For local dwarf spheroidal galaxies \citet{gilmore:07} derived the
central characteristic DM density (averaged over a volume of radius
10\,pc) within a cusped model
\mbox{$\sim\!60$}\,$\rmn{M_{\sun}pc^{-3}}$. For our models, the
corresponding values are 33 for the NFW profile and 16 for
the Einasto profile.
For the cored Burkert profile, this value is
\mbox{$0.057\,\rmn{M_{\sun}pc^{-3}}$}. It is interesting to note
that the central characteristic density of the DM for cuspy profiles
is nearly independent of the galactic mass. This value is close to
the upper limit of the central DM  density range of nine local
galaxies (1--4)$\cdot 10^{-24}\,\rmn{g\,cm^{-3}} =$
(0.015--0.059)$\,\rmn{M_{\sun}pc^{-3}}$ derived within the cored DM
models by \citet{borriello:01}. For four galaxies at mean redshifts
$\langle z\rangle \simeq 0.9$, \citet{tamm:05} derived the central
DM density (0.012--0.028)$\,\rmn{M_{\sun}pc^{-3}}$. This lower value
at higher redshift may hint to the cosmological evolution of DM
densities.

\section{Conclusions}\label{sec:7}

In the present paper we have derived the density distribution of DM
in a most well-observed nearby disc galaxy, the Andromeda galaxy. In
the first part of the study (Paper~I), we construct the
photometrical model of M\,31 stellar populations on the basis of
surface brightness profiles in {\it U\/}, {\it B\/}, {\it V\/}, {\it
R\/}, {\it I\/} and {\it L\/} colours. The derived photometrical
model together with metallicities is used in chemical evolution
models to calculate the $M/L$-ratios of components. In the second
part of the study (the present paper), we construct a consistent
mass distribution model of M\,31. Calculating the rotation
velocities and velocity dispersions of visible matter with the help
of the dynamical model we found the amount of DM which must be added
to reach an agreement with the observed rotation and dispersion
data.

We conclude that within uncertainties the Moore, the Burkert,
the NFW and the Einasto distributions fit with observational
limits at nearly all distances (Fig.~\ref{dm density}). However, it
is also seen that in central regions, the  Moore and the
Burkert profiles lie rather close to the upper and lower limits of
uncertainties. The Burkert law does not fit within the lower limit for
the mass of M\,31, set by \citet{brunthaler:07}, either. The NFW and the
Einasto density distributions give the best fit.

$\rmn{\Lambda CDM}$ hierarchical clustering theory is in
agreement with dark matter density distribution in a well studied
disc galaxy M\,31. Predictions of this theory also agree with
observed decrease of disc sizes with redshift \citep{bouwens:02,ferguson:04,tamm:06}.

\section*{Acknowledgments}
The paper has been improved according to the comments of the anonymous referee.
We thank Dr.~J.~Pelt for suggesting the bootstrap method for
averaging the observational data. We thank Dr.~P.~Salucci and
Dr.~C.~Frigerio~Martins for pointing out the problem with the NFW
concentration parameter. We thank Dr.~A.W.~Graham for the suggestion
to use the Einasto model instead of the \citet{navarro:04} model.
We acknowledge the financial support from the Estonian Science
Foundation {grants 6106, 7115, 7146} and project SF0060067s08. All
figures were made with the GNUPLOT plotting utility.


\label{lastpage}
\end{document}